\documentclass[a4paper,11pt]{article}
\usepackage{graphicx} 
\usepackage[applemac]{inputenc}

\author{M. Cosentino Lagomarsino\footnote{ FOM Institute for Atomic and
Molecular Physics (AMOLF), Kruislaan 407, 1098 SJ Amsterdam, The
Netherlands;
Tel. +31 - (0)20 - 6081234 ; fax +31 - (0)20 - 6684106 
; e-mail address cosentino-lagomarsino@amolf.nl },
B. Bassetti\footnote{Università degli Studi di Milano, Dip. Fisica, Via
Celoria 16, 20100 Milano, Italy; 
Tel. +39 - (0)2 - 58357477 ; fax +39 - (0)2 - 58357480 
; e-mail address bassetti@mi.infn.it },
P. Jona \footnote{Politecnico di Milano, Dip. Fisica, Pza Leonardo Da
Vinci 32, 20100 Milano, Italy;
Tel. +39 - (0)2 - 23996133 ; fax +39 - (0)2 - 23996126
; e-mail address patrizia.jona@polimi.it}}

\title{{\it Rowers} Coupled Hydrodynamically. Modeling Possible
                    Mechanisms for the Cooperation of Cilia} 

\begin{document}
   
\maketitle

{\large 
\begin{center}
\end{center}
}

\begin{abstract}
  We introduce a model system of stochastic entities, called {\it
    rowers} which include some essentials of the behavior of real
  cilia. We introduce and discuss the problem of symmetry breaking for
  these objects and its connection with the onset of macroscopic,
  directed flow in the fluid. We perform a mean field-like calculation
  showing that hydrodynamic interaction may provide for the symmetry
  breaking mechanism and the onset of fluid flow. Finally, we discuss
  the problem of the metachronal wave in a stochastic context through
  an analytical calculation based on a path integral representation
  of our model equation.
\end{abstract}
\newpage 
\section{Introduction and overview}\label{sec.1}

We set  up a model system  of entities, called {\it  rowers}, that may
organize spontaneously  breaking left-right symmetry of the motion and
  give rise to an ordered  macroscopic flow and beating patterns. 
Rowers are active,
stochastic elements that ''live'' in  a fluid with low Reynolds number
and are  capable to exert influence  on each other by means of
hydrodynamic coupling.

The motivation for this analysis  comes from the motion of cilia, long
and thin extroflections of the  eucaryotic cell membrane that are able
to generate  motion \cite{book} \cite{Sleigh}.  The cilia  are used by
the  cell for  self-propulsion or  to stir  the surrounding  fluid.  A
cilium  has   an  internal  structure  (the   axoneme)  containing  an
arrangement of microtubule doublets  attached to a basal body anchored
to the cell membrane. A  complex, symmetric net of protein bridges and
links among the doublets gives the whole structure elastic properties.
Biochemical  reactions  at the  level  of such  proteins
represent  the  energy source  for  the ciliary motion,
a cyclic beating composed  of two  phases: the  {\it effective
stroke} -- which is active in  propulsion or in fluid transport -- and
the  {\it  recovery  stroke},  which  is  passive.   A  ciliated  cell
generally has a field of hundreds of cilia which beat in a coordinated
manner, setting up wave-like  time dependent patterns. This phenomenon
is referred to as {\it metachronism}.

In the framework of a purely mechanical description, 
the physics of ciliary motion involves the balance of
hydrodynamic and elasticity; the forced nature of the system is modeled
through the elastic constitutive equations, which contain the 
{\it engine} supplying energy in a deterministic way.

In this paper we adopt a stochastic approach to ciliary motion and its 
associated macroscopic fluid flux.
We analyze some physical assumptions which are 
required as necessary conditions for the existence of a macroscopic net 
flow in the surrounding fluid and for the onset of coordination, or 
metachronism. \\ 
Our spirit is to approach the system from the point of view of 
Statistical Mechanics, looking at macroscopic effects and keeping as 
few as possible the number of relevant variables. 
As our model is designed to be studied as much as possible with
analytical tools, we do not model in detail the internal features of
the single object- the rower - which inherits from the real cilium the
only peculiarity of undergoing a two-phase motion.  We consider
hydrodynamic interaction and energy supply, which is switched on and
off at times controlled by a stochastic process, and possibly
correlated to the configuration of the object.

We think that some of the questions pointed out in this study,
may be of general interest --independently of the problem of ciliary motion --
for the Statistical Mechanics in far from equilibrium systems.
 
Looking at the literature about  cilia and flagella in a viscous fluid
one realizes that such studies are mainly mechanics--oriented, the
stochastic  aspects  of  the  phenomena being,  in  general,  disregarded.

Early works on these subjects  date since 1955 with the early
model by Gray and Hancock  (one  cilium in  a  fluid) \cite{GH55}
and with the study by Machin (1963 -  locally contractile flagellum)  \cite{Machin}.  
Later,  mathematical models  for simulating
the motion of cilium (and  flagellum) have been developed, 

we refer to the works by Brokaw {\it et.al.}
\cite{Br72} 
\cite{Br76}  
\cite{HB78}  
\cite{Li76}  
\cite{JB79}, 
(curvature controlled models)
and by Murase {\it et.al.} 
\cite{MS86} 
\cite{MHB89}
\cite{Mu90} 
\cite{Mu91}
\cite{Mu90} 
(excitable dynein models and directional  mechano-sensitivity  of  cilium  
as a possible mechanism for the onset of  metachronal waves  
{\it via} hydrodynamic interaction \cite{Mu90}) .

The description of 
metachronism,  was  addressed  to  model  multicilia
dynamical configurations in a suitable  way to generate Stokes flow in
the surrounding fluid  (Liron, Blake {\it et al.} 
\cite{LS72} \cite{Bl72} \cite{LM76a});
the  problem  of  fluid   transport  was investigated for  different
geometries  of fluid  confinement
\cite{Bl73}
\cite{LM76b}  
\cite{Li78} 
\cite{LS78}
\cite{BLA82}.
In  1984 Liron in his  work \cite{Li84}, based  on the 
discrete   cilia  approach  \cite{LS78},  
described fluid  transport  by cilia, assuming
metachronal coordination between ciliary  moves (propagating wave) 
and periodicity conditions.\\
More recently,  Gueron {\it et al.}  proposed 
a model which accounts for multicilia  hydrodynamical interactions  
\cite{GL92}  
\cite{GL93} \cite{GLL97} \cite{GL97}  
and energetics 
\cite{gueron1}.  
The  dependence   of  the  metachronal  wave   on  observable  ciliary
parameters \cite{gheber2}
 -- and the  effect of varying fluid viscosity \cite{gheber}
have  been studied  by  Priel  et  al.   
They  also  proposed  a  model  involving
hydrodynamically  coupled oscillators \cite{GP89}  
\cite{PT90}
\cite{P.et.al.}.

In building up the model, we represent the system in terms of
two-phase hydrodynamically coupled oscillators.  The nonequilibrium
drive (active motion) is realized as stochastic transitions between
two internal states, acting together with gaussian thermal
fluctuations.

In this paper we are interested in focusing two crucial problems. 
The first is symmetry breaking.  

If the internal (mechanical) engine which generates motion is removed
the cilium is free of moving in a cone with cylindrical symmetry, O(2)
in 3D with fixed basis.  However, cilia and flagella are observed to
perform planar motion.
In the case of sea-urchin sperm cell flagella the plane of
motion can be imposed by external perturbations \cite{Shing}.

The internal couple of microtubules in the axoneme -- may give an
explanation to the breaking of the O(2) symmetry. 

The problem whether, once moving in a plane, there is any preference
for right or left-directed effective stroke is, to our knowledge,
open.  It is invaluable by purely anatomic reasons of the individual,
that is, there is no \emph{structural} symmetry breaking. Flagella,
for example, are observed to beat symmetrically.

In \emph{paramecia} this left right symmetry is broken in connection
with an intrinsically oriented structure of the whole cell cortex (the
so-called kineties), which is absent in ciliated epithelial cells
\cite{sonneborn}.

Therefore real cilia need a symmetry-breaking mechanism to push the
fluid in a directed way. This could be due to regulatory processes
that establish a cortical anisotropy followed by the cilia.  With our
simplified model we show that hydrodynamic interactions in collective
motions of rowers could be enough to realize this symmetry breaking.
This situation is interesting from the point of view of statistical
mechanics, because a local, a priori isotropic release of energy is
transformed in a self organized way in a macroscopically relevant
collective state with a well-defined directionality (see also
\cite{reimann}).

The second problem is the physical source of metachronism.  With our
model we show that metachronism is not a wave phenomenon of the
traditional kind, but, instead, a phenomenon of statistical nature.
In fact, it can not be sustained by oscillations around the ground
state of the system in thermodynamical equilibrium, but it is more
understandable as a time-dependent pattern created by the
counteracting active beating and dissipative processes.

In section \ref{sec.2}. we present the general features of the model.
The elementary component is a rather abstract object (the rower).  It
includes a few observed features of real cilia, mainly the distinction
between effective stroke and recovery stroke.  Our rowers are one
dimensional but not intrinsically oriented, that is, they have
left-right symmetry.  The observed two-phase beating is represented as
the motion of a particle in two different potentials -- active and
passive -- alternatively switched on and off by a two-state 
stochastic process.
These potentials are analogues of states of a filament in which 
the collective action of the dyneins
determines two different minimum energy curvatures. 

In section  \ref{sec.3}. we show how to compute averaged  quantities 
for  the single rower, of which the most  interesting is the current 
that it generates as a function of the external velocity of the fluid.\\
We then use this result, together with known techniques of
fluid mechanics,  to compute the  self-consistent velocity field  in a
low Reynolds number  Stokes fluid with an array  of rowers as velocity
sources.   This  can be  taken  as  a  demonstration that  cooperative
effects  arising  from  the  hydrodynamic  interaction  may  make  our
stochastic  rowers   spontaneously  break   symmetry  and  set   up  a
macroscopic flux. This result is obtained in a mean-field like picture,
and is independent of a more-refined investigation on the collective 
motion of rowers.

In the last section, we discuss the premises for the onset of
metachronal waves, defined as spatio-temporal anisotropic ground
states of the system.  Pointed out the role of hydrodynamic
interactions in models that include thermal noise, we proceed
employing a path integral representation of our model equations to
obtain information on the most probable history (path in the
configurational space) and first
excitations.\\
We analyze in more detail the case in which 
any feedback mechanism of position and internal state is
avoided. The cyclic nature of ciliary motion (i.e. the fact that its
intrinsic time scale is larger than relaxation times) is taken
explicitly into account.  We prove that the onset of metachronal waves
is compatible with such a system, but it is frustrated by the presence
of random solutions with the same statistical weight. 

This means that, if there is no exchange of signals between cilia,
(nether chemical nor mechanical) the onset of metachronism due to
hydrodynamic interaction is not relevant. However, a nearest neighbor
coupling between the internal state of the rowers is enough to
stabilize this wave-like solution.  

We also argue that if the  transition probabilities between states are
coupled with the configuration of the rower -- the coupling may be
realized for example as a stochastic version of the ``geometric
switch'' introduced by Gueron {\it et al.} \cite {gueron2} --
the problem of metachronism becomes formally analogous to the problem
of modulated phases in membranes with defects \cite{leiblerfasi}.

\section{The Rower}\label{sec.2}

A  rower is the  elementary tile of  our model, and is  designed to
contain some of the essentials of the cilium.\\
Our  starting point  is  the  observation that  a  single cilium  beat
pattern  can be  divided into  two phases  \cite{Sleigh}.   During the
\emph{effective stroke}  it   moves  almost   as  a   straight  rod,
transversally to the fluid, while during the \emph{recovery stroke} it
glides back softly, in a tangential motion.  Thus the effective stroke
is associated with  high viscous load and actually  propels the fluid,
whereas the recovery stroke brings  back the cilium to its equilibrium
position minimizing the viscous resistance.

A rower is characterized by two degrees of freedom, a continuous one
corresponding to its position (the cilium center of mass, for
example), and a discrete one corresponding to its internal state. The
rower is alternatively subject to two different potentials and viscous
loads, with different hydrodynamical characterization.  
In state \( 1 \) (recovery stroke) the viscous coefficient is
low and the particle ``sees'' a concave potential \( V_{1} \), while
in state \( 2 \) (effective stroke) the viscous coefficient is high
and the potential \( V_{2} \) looks like a mexican hat.  The
transitions between the two states are stochastic.  The switching
between two potentials makes our rower an active element, and is the
analogue of the active component of the force in the mechanical models
of the cilium.  The two different viscous loads mimic the behavior of
a slender body moving transversely or tangentially in the fluid.

We consider a one-dimensional rower, that is, the rower breaks
rotational but not left-right symmetry.  This is different from most
models of ciliary activity found in the literature, which treat cilia
as structurally asymmetric objects.  In both states one has to take
into account the external drive due to the velocity of the fluid.  At
one-body level, the velocity of the external fluid enters as a linear
bias the active and passive potentials; with such a bias the
left-right symmetry will be broken.

Making the additional hypothesis that we are in the overdamped regime
(low Reynolds number) \cite{purcell}, we can use the following
Langevin equation to describe the dynamics of each rower (see
\cite{astumian} for a similar equation in a different context)
\begin{equation}
     \dot{x} = v - \frac{1}{\gamma_{\sigma} } \frac{\partial
V_{\sigma}(x)}{\partial x} + \xi_{\sigma} \label{eq:1}
\end{equation} 
where \( \sigma(t) \) may take values in \( \{ -1,1 \} \) and is the
stochastic process describing the switching between phases. \\ 
The simplest choice for this noise is a random
telegraph process, with Poisson distributed jumps. 
Alternatively, one could allow the transitions between the two states
to depend on the configuration of the rower.
For example one could require  the transition probability to increase at the
two ends of the rowing oscillations , with a  mechanism which is the analogue
for our rower of the ``geometric switch'' introduced in\cite{gueron2}.\\
We will stick for simplicity of exposition to the first case
throughout this section and the next, as the results do not change in
substance from the point of view of a ``mean field''-like description.
In section \ref{sec.4} we will distinguish between the two mechanisms.

In eq. (1) \(\xi_{\sigma} \) is a gaussian white noise with zero average and
correlation
\begin{displaymath}
    \langle \xi_{\sigma}(t) \xi_{\sigma}(t') \rangle =
\frac{2T}{\gamma_{\sigma}} \delta(t - t')
\end{displaymath} 
In the same equation \( v \) is the component of the surrounding fluid velocity along the
direction of our one-dimensional rower.\\ 
In the dynamics described by equation \ref{eq:1} we have eliminated
two ``fast'' modes with characteristic times \( \tau_{+,\sigma} =
\frac{m}{\gamma_{\sigma}}\), where \( m \) is the mass of the
particle. Thus, we are left with the ``slow'' modes with relaxation
times \( \tau_{-,\sigma} \simeq
\frac{\gamma_{\sigma}}{\kappa_{\sigma}} \), where \( \kappa_{\sigma}
\) is the curvature of potential \( V_{\sigma} \) at its minimum(s).
In order for the model to represent effectively the movement of a
rower, the average time between two stochastic transitions must be
greater than the characteristic times \( \tau_{-,\sigma} \).  This
implies in the first place that the system has to be far from
criticality (see \cite{ma}).

Of course our rowers overlook many details of the mechanism of
contraction of real cilia. In the first place they are not filaments
but ''points''. On the other hand we do not want to focus on the detailed
modeling of real cilia but instead on the organization through
hydrodynamic coupling, and they are designed to this aim. In fact, the
statistical mechanics of an internally driven filament  is quite a
difficult subject (one object has infinitely many degrees of freedom),
while our rowers turn out to be much milder.


\section{Symmetry breaking and onset of macroscopic flux}\label{sec.3}

\subsection{Single rower}

It is convenient to pass from equation \ref{eq:1} to the
Fokker-Planck description, that deals with the distribution functions
\( P_{\sigma}(x,t) \) for the probability to find the rower with
position \( x \) at time \( t \) in state \( \sigma \).  These
distributions obey the equations
\begin{equation}
    \left\{ \begin{array}{c} \partial_{t} P_{1}(x,t) = -\partial_{x}
    J_{1}(x,t) -\omega_{1} P_{1}(x,t) + \omega_{2} P_{2}(x,t) \\ \\
    \partial_{t} P_{2}(x,t) = -\partial_{x} J_{2}(x,t) +\omega_{1}
    P_{1}(x,t) - \omega_{2} P_{2}(x,t) \end{array} \right.
\label{eq:2}
\end{equation}

with probability currents 
\begin{displaymath}
    J_{\sigma} = vP_{\sigma} - \frac{1}{\gamma_{\sigma}}\frac{\partial
    V_{\sigma}}{\partial x} P_{\sigma} -
    \frac{T}{\gamma_{\sigma}}\frac{\partial P_{\sigma}}{\partial x}
\end{displaymath}
where \( \omega_{\sigma} \) is the probability of transition from the
state labeled by \( \sigma \) to the other (these quantities would
depend on \( x \) if one chooses a configuration-dependent case).  
Typical values for these quantities are around \( 60 \:s^{-1} \). \\
The motion of the rower looks like an oscillation between
the bottoms of the two potentials, and each phase of the cycle has a
mean duration of \( \tau_{i} = \omega^{-1}_{i} \).

Equations \ref{eq:1} and \ref{eq:2} resemble formally those of a
two-state thermal ratchet model (e.g. \cite{astumian}).  
Actually, in our model the physical situation is quite different.  
In fact, if we consider the stationary Fokker-Planck equation for  
the  sum of probability currents \( J_{tot} = J_{1} + J_{2} \), because of the 
absence of periodic boundary condition,  
 \(  \partial_{x} J_{tot} = 0 \) implies
that \(J_{tot} \)  is zero. That is, a global flow of probability
is not possible: the rower can't just run away.\\
One should not worry about the fact that the overall net current is zero, 
because the two strokes of the rower produce much different perturbations 
in the surrounding fluid, thanks to the differences in viscosity. 
The problem becomes then if the two probability currents \( J_i (x) \), 
which have opposite sign, are nonzero. 

\begin{figure}[tbp]
\begin{center}
\input{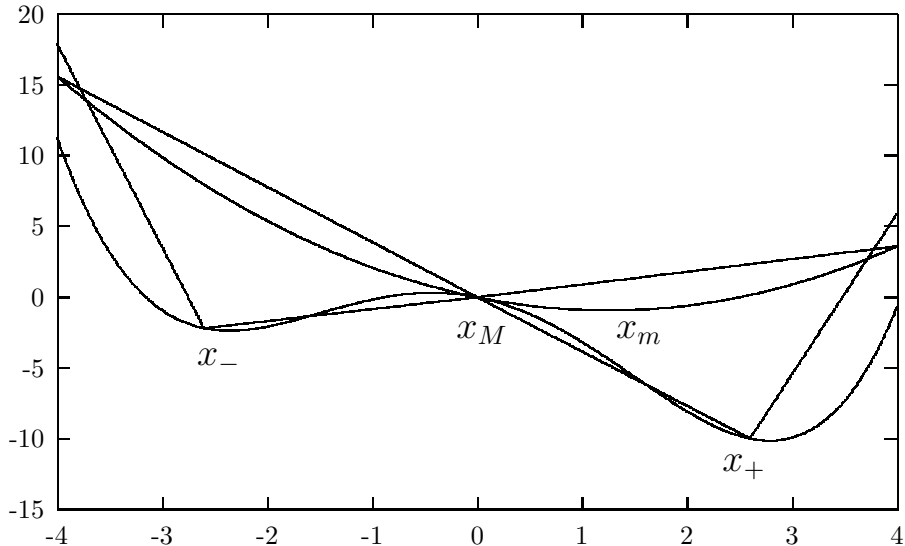}
\end{center}
\caption{Model potentials with incorporated external fluid velocity. 
For the passive phase \( V_{1}(x)=\frac
{a}{2}x^2-v x \) ; for the active phase \(V_{2}(x)=\frac
{b}{4}x^2(x^2-x_0^2)-vx\). Here rescaled parameters are plotted 
\( a= 1.2; b= 0.8; x_0 =3.7 ; v= 1.5; \gamma_1= 1.0; \gamma_2= 1.5 \). 
Piece-wise linear potentials are also drawn. 
Realistic parameters are: 
\( x_0 \sim 5 \mu \textrm{m}; v \sim 10^{-5} \textrm{ to } 10^{-4} \mu \textrm{m/s }; 
\gamma_2 \simeq 1.5
\gamma_1, \gamma_1 \sim 10^{-3} \textrm{N s/m}; a \sim 10^{-6} N/m
\).}
\label{fig:1} 
\end{figure}

Let us compute the mean stationary value of the velocity during
the effective stroke.  We can write the average active current as

\begin{equation}
   I \equiv  I_2 = - I_1 
        \label{eq:3}
\end{equation}

with

\begin{displaymath}
   I_i  = \int J_{i}(x) \textrm{dx} 
\end{displaymath}
With a little manipulation of equation \ref{eq:2} it is easy to
obtain a third order differential equation for \( P_{1} \) (or \( P_{2}
\)) that, once solved, allows to compute explicitly the above
averages.  We have been able to solve analytically this equation
by transfer matrix method
in the case of piece-wise linear potentials (figure \ref{fig:1}).
For more general cases, typically that of a quadric \( V_{1} \) and a
quartic \( V_{2} \) (figure \ref{fig:1}), we have resorted to solve
numerically equation \ref{eq:2} and look at its long-time behavior.

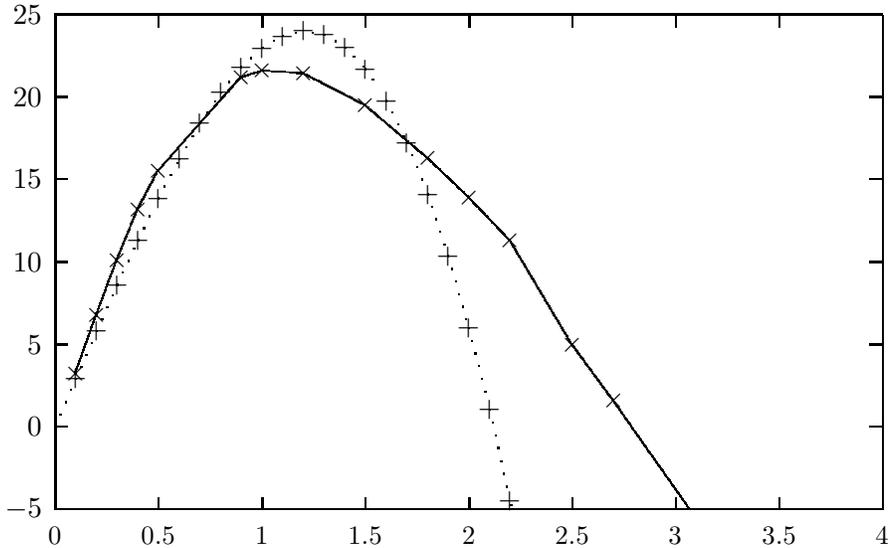
\begin{figure}[h]
\begin{center}
\setlength{\unitlength}{0.240900pt}
\ifx\plotpoint\undefined\newsavebox{\plotpoint}\fi
\begin{picture}(1500,900)(0,0)
\font\gnuplot=cmr10 at 10pt
\gnuplot
\sbox{\plotpoint}{\rule[-0.200pt]{0.400pt}{0.400pt}}%
\put(140.0,82.0){\rule[-0.200pt]{4.818pt}{0.400pt}}
\put(120,82){\makebox(0,0)[r]{$-5$}}
\put(1419.0,82.0){\rule[-0.200pt]{4.818pt}{0.400pt}}
\put(140.0,212.0){\rule[-0.200pt]{4.818pt}{0.400pt}}
\put(120,212){\makebox(0,0)[r]{$0$}}
\put(1419.0,212.0){\rule[-0.200pt]{4.818pt}{0.400pt}}
\put(140.0,341.0){\rule[-0.200pt]{4.818pt}{0.400pt}}
\put(120,341){\makebox(0,0)[r]{$5$}}
\put(1419.0,341.0){\rule[-0.200pt]{4.818pt}{0.400pt}}
\put(140.0,471.0){\rule[-0.200pt]{4.818pt}{0.400pt}}
\put(120,471){\makebox(0,0)[r]{$10$}}
\put(1419.0,471.0){\rule[-0.200pt]{4.818pt}{0.400pt}}
\put(140.0,601.0){\rule[-0.200pt]{4.818pt}{0.400pt}}
\put(120,601){\makebox(0,0)[r]{$15$}}
\put(1419.0,601.0){\rule[-0.200pt]{4.818pt}{0.400pt}}
\put(140.0,730.0){\rule[-0.200pt]{4.818pt}{0.400pt}}
\put(120,730){\makebox(0,0)[r]{$20$}}
\put(1419.0,730.0){\rule[-0.200pt]{4.818pt}{0.400pt}}
\put(140.0,860.0){\rule[-0.200pt]{4.818pt}{0.400pt}}
\put(120,860){\makebox(0,0)[r]{$25$}}
\put(1419.0,860.0){\rule[-0.200pt]{4.818pt}{0.400pt}}
\put(140.0,82.0){\rule[-0.200pt]{0.400pt}{4.818pt}}
\put(140,41){\makebox(0,0){0}}
\put(140.0,840.0){\rule[-0.200pt]{0.400pt}{4.818pt}}
\put(302.0,82.0){\rule[-0.200pt]{0.400pt}{4.818pt}}
\put(302,41){\makebox(0,0){0.5}}
\put(302.0,840.0){\rule[-0.200pt]{0.400pt}{4.818pt}}
\put(465.0,82.0){\rule[-0.200pt]{0.400pt}{4.818pt}}
\put(465,41){\makebox(0,0){1}}
\put(465.0,840.0){\rule[-0.200pt]{0.400pt}{4.818pt}}
\put(627.0,82.0){\rule[-0.200pt]{0.400pt}{4.818pt}}
\put(627,41){\makebox(0,0){1.5}}
\put(627.0,840.0){\rule[-0.200pt]{0.400pt}{4.818pt}}
\put(790.0,82.0){\rule[-0.200pt]{0.400pt}{4.818pt}}
\put(790,41){\makebox(0,0){2}}
\put(790.0,840.0){\rule[-0.200pt]{0.400pt}{4.818pt}}
\put(952.0,82.0){\rule[-0.200pt]{0.400pt}{4.818pt}}
\put(952,41){\makebox(0,0){2.5}}
\put(952.0,840.0){\rule[-0.200pt]{0.400pt}{4.818pt}}
\put(1114.0,82.0){\rule[-0.200pt]{0.400pt}{4.818pt}}
\put(1114,41){\makebox(0,0){3}}
\put(1114.0,840.0){\rule[-0.200pt]{0.400pt}{4.818pt}}
\put(1277.0,82.0){\rule[-0.200pt]{0.400pt}{4.818pt}}
\put(1277,41){\makebox(0,0){3.5}}
\put(1277.0,840.0){\rule[-0.200pt]{0.400pt}{4.818pt}}
\put(1439.0,82.0){\rule[-0.200pt]{0.400pt}{4.818pt}}
\put(1439,41){\makebox(0,0){4}}
\put(1439.0,840.0){\rule[-0.200pt]{0.400pt}{4.818pt}}
\put(140.0,82.0){\rule[-0.200pt]{312.929pt}{0.400pt}}
\put(1439.0,82.0){\rule[-0.200pt]{0.400pt}{187.420pt}}
\put(140.0,860.0){\rule[-0.200pt]{312.929pt}{0.400pt}}
\put(140.0,82.0){\rule[-0.200pt]{0.400pt}{187.420pt}}
\put(172,296){\usebox{\plotpoint}}
\multiput(172.58,296.00)(0.497,1.403){63}{\rule{0.120pt}{1.215pt}}
\multiput(171.17,296.00)(33.000,89.478){2}{\rule{0.400pt}{0.608pt}}
\multiput(205.58,388.00)(0.497,1.352){61}{\rule{0.120pt}{1.175pt}}
\multiput(204.17,388.00)(32.000,83.561){2}{\rule{0.400pt}{0.588pt}}
\multiput(237.58,474.00)(0.497,1.219){63}{\rule{0.120pt}{1.070pt}}
\multiput(236.17,474.00)(33.000,77.780){2}{\rule{0.400pt}{0.535pt}}
\multiput(270.58,554.00)(0.497,0.941){61}{\rule{0.120pt}{0.850pt}}
\multiput(269.17,554.00)(32.000,58.236){2}{\rule{0.400pt}{0.425pt}}
\multiput(302.58,614.00)(0.499,0.565){257}{\rule{0.120pt}{0.552pt}}
\multiput(301.17,614.00)(130.000,145.854){2}{\rule{0.400pt}{0.276pt}}
\multiput(432.00,761.58)(1.534,0.492){19}{\rule{1.300pt}{0.118pt}}
\multiput(432.00,760.17)(30.302,11.000){2}{\rule{0.650pt}{0.400pt}}
\multiput(465.00,770.93)(7.167,-0.477){7}{\rule{5.300pt}{0.115pt}}
\multiput(465.00,771.17)(54.000,-5.000){2}{\rule{2.650pt}{0.400pt}}
\multiput(530.00,765.92)(0.973,-0.498){97}{\rule{0.876pt}{0.120pt}}
\multiput(530.00,766.17)(95.182,-50.000){2}{\rule{0.438pt}{0.400pt}}
\multiput(627.00,715.92)(0.590,-0.499){163}{\rule{0.572pt}{0.120pt}}
\multiput(627.00,716.17)(96.812,-83.000){2}{\rule{0.286pt}{0.400pt}}
\multiput(725.00,632.92)(0.524,-0.499){121}{\rule{0.519pt}{0.120pt}}
\multiput(725.00,633.17)(63.922,-62.000){2}{\rule{0.260pt}{0.400pt}}
\multiput(790.58,569.85)(0.499,-0.523){125}{\rule{0.120pt}{0.519pt}}
\multiput(789.17,570.92)(64.000,-65.923){2}{\rule{0.400pt}{0.259pt}}
\multiput(854.58,501.81)(0.499,-0.838){193}{\rule{0.120pt}{0.769pt}}
\multiput(853.17,503.40)(98.000,-162.403){2}{\rule{0.400pt}{0.385pt}}
\multiput(952.58,338.34)(0.499,-0.677){127}{\rule{0.120pt}{0.642pt}}
\multiput(951.17,339.67)(65.000,-86.668){2}{\rule{0.400pt}{0.321pt}}
\multiput(1017.58,250.22)(0.499,-0.713){237}{\rule{0.120pt}{0.670pt}}
\multiput(1016.17,251.61)(120.000,-169.609){2}{\rule{0.400pt}{0.335pt}}
\multiput(140,212)(8.145,19.090){4}{\usebox{\plotpoint}}
\multiput(172,287)(8.359,18.998){4}{\usebox{\plotpoint}}
\multiput(205,362)(8.333,19.009){4}{\usebox{\plotpoint}}
\multiput(237,435)(8.851,18.774){4}{\usebox{\plotpoint}}
\multiput(270,505)(9.055,18.676){3}{\usebox{\plotpoint}}
\multiput(302,571)(9.752,18.322){4}{\usebox{\plotpoint}}
\multiput(335,633)(10.298,18.021){3}{\usebox{\plotpoint}}
\multiput(367,689)(11.759,17.103){3}{\usebox{\plotpoint}}
\multiput(400,737)(12.966,16.207){2}{\usebox{\plotpoint}}
\multiput(432,777)(15.358,13.962){2}{\usebox{\plotpoint}}
\multiput(465,807)(17.847,10.596){2}{\usebox{\plotpoint}}
\multiput(497,826)(20.171,4.890){2}{\usebox{\plotpoint}}
\put(543.81,831.41){\usebox{\plotpoint}}
\multiput(562,828)(17.750,-10.758){2}{\usebox{\plotpoint}}
\multiput(595,808)(14.225,-15.114){3}{\usebox{\plotpoint}}
\multiput(627,774)(11.433,-17.323){2}{\usebox{\plotpoint}}
\multiput(660,724)(9.055,-18.676){4}{\usebox{\plotpoint}}
\multiput(692,658)(7.831,-19.222){4}{\usebox{\plotpoint}}
\multiput(725,577)(6.502,-19.711){5}{\usebox{\plotpoint}}
\multiput(757,480)(5.655,-19.970){6}{\usebox{\plotpoint}}
\multiput(789,367)(5.182,-20.098){6}{\usebox{\plotpoint}}
\multiput(822,239)(4.532,-20.255){7}{\usebox{\plotpoint}}
\put(855.55,88.76){\usebox{\plotpoint}}
\put(857,82){\usebox{\plotpoint}}
\sbox{\plotpoint}{\rule[-0.500pt]{1.000pt}{1.000pt}}%
\put(172,296){\makebox(0,0){$\times$}}
\put(205,388){\makebox(0,0){$\times$}}
\put(237,474){\makebox(0,0){$\times$}}
\put(270,554){\makebox(0,0){$\times$}}
\put(302,614){\makebox(0,0){$\times$}}
\put(432,761){\makebox(0,0){$\times$}}
\put(465,772){\makebox(0,0){$\times$}}
\put(530,767){\makebox(0,0){$\times$}}
\put(627,717){\makebox(0,0){$\times$}}
\put(725,634){\makebox(0,0){$\times$}}
\put(790,572){\makebox(0,0){$\times$}}
\put(854,505){\makebox(0,0){$\times$}}
\put(952,341){\makebox(0,0){$\times$}}
\put(1017,253){\makebox(0,0){$\times$}}
\sbox{\plotpoint}{\rule[-0.200pt]{0.400pt}{0.400pt}}%
\put(172,287){\makebox(0,0){$+$}}
\put(205,362){\makebox(0,0){$+$}}
\put(237,435){\makebox(0,0){$+$}}
\put(270,505){\makebox(0,0){$+$}}
\put(302,571){\makebox(0,0){$+$}}
\put(335,633){\makebox(0,0){$+$}}
\put(367,689){\makebox(0,0){$+$}}
\put(400,737){\makebox(0,0){$+$}}
\put(432,777){\makebox(0,0){$+$}}
\put(465,807){\makebox(0,0){$+$}}
\put(497,826){\makebox(0,0){$+$}}
\put(530,834){\makebox(0,0){$+$}}
\put(562,828){\makebox(0,0){$+$}}
\put(595,808){\makebox(0,0){$+$}}
\put(627,774){\makebox(0,0){$+$}}
\put(660,724){\makebox(0,0){$+$}}
\put(692,658){\makebox(0,0){$+$}}
\put(725,577){\makebox(0,0){$+$}}
\put(757,480){\makebox(0,0){$+$}}
\put(789,367){\makebox(0,0){$+$}}
\put(822,239){\makebox(0,0){$+$}}
\put(854,96){\makebox(0,0){$+$}}
\end{picture}
\end{center}
\caption{Active current \( I(v) \) as a function of the surrounding fluid
velocity from calculation with the potentials of fig.1 (points with dotted 
line for piece-wise linear potential).}
\label{fig:2} 
\end{figure}

From simulations and calculations it's clear that, when \( v = 0 \), \(
I_1 = I_2 = 0 \) and there is no biased pumping. For \( v \not = 0 \) instead
the average currents are finite and sustain fluid flow.
In figure \ref{fig:2} we show the computed average active current \( I(v)
\) as a function of the surrounding fluid velocity; and it is
nonzero for \( v \ne 0 \) --  negative values for \(I(v) \) are just an 
artifact, because by increasing \(v\) the minimum of \( V_1 \) passes the 
right minimum of \(V_2\).

We now give a heuristic argument that, for low temperature and
driving velocity, justifies this behavior.  In these conditions, the
process is well approximated by a sequence of jumps between the
minimums of the two potentials.  These jumps are unbiased as long as
there is no driving velocity, since the left-right symmetry is not
broken.  The presence of the linear term induces a bias in the jump
probability , so that from the minimum \( x_{m}(v)
\) of \( V_{1} \) (figure \ref{fig:1}) the rower has a probability 
of \( 1/2 + \pi(v) \) to fall into \( x_{+} \) and \( 1/2 - \pi(v) \) 
to fall into \( x_{-} \).
Thus we can estimate the average current as \( \frac{2\pi(v)}{p} \),
where p is the period of the rowing cycle.\\
If we make the further assumption that in the recovery phase the
probability distribution for the position of the rower relaxes to a
gaussian centered in \( x_{m}(v) \), with width \( \frac{T}{a} \), it
is easy to see that
\begin{displaymath}
    \frac{1}{2} + \pi(v) =\sqrt{ \frac{T}{a}} \textrm{erf}(x_{m}(v) - x_{M}(v))
\end{displaymath}
which, substituting, gives the same qualitative behavior for the average 
current as that shown in figure \ref{fig:2}.  

Thus, for nonzero external velocity field \( v \), \( I(v)\ne 0 \),
and the rower breaks left-right symmetry. In this situation, the average
{\it excess} Stokes force exerted by the rower  on the surrounding fluid, 
is  
\begin{displaymath}
        F_S = (\gamma_{2} - \gamma_{1}) I(v)
\end{displaymath}
and there is biased pumping of the fluid as long as the two 
viscosities are different. In the above expression, the information on
the transition times is contained in \(I\). 
The dependence of the force on the fluid velocity is reminiscent of
the \emph{mechano-sensitivity} found by Murase in \cite{Mu90}.

\subsection{Array of rowers}

In the case of a planar array of \( N \) interacting rowers beating in
the same direction and arranged on a lattice (which can also be random), 
we have \( N \) equations that look like equation \ref{eq:1}, 
but the surrounding fluid velocity term \( v \) has to 
be substituted by the contribution of all the other rowers
through hydrodynamic interactions.  That is, we have a sum of all the
other particles velocities to which we apply the mobility matrix, 
which we take as \(\bar{H}_{II} = {\mathbf 1} \), \( \bar{H}_{IJ} =  
\mathbf{H}(\mathbf{r}_{IJ}) \), where 
\( \mathbf{H}(\mathbf{r}) = \frac{1}{8\pi \eta r} ({\mathbf 1} + \mathbf{\hat{r}} 
\mathbf{\hat{r}}) \) is the Oseen tensor,
which has a dependence on the inverse distance (see \cite{Doi} p.68).
Then we write (\( I \) and \( J \) label lattice sites and 
\(\mathbf{ \hat {d}}\) is the unit vector directed along the rower's motion):

\begin{displaymath}
  \mathbf{v}_I= \sum_{J \ne I}\bar{H}_{IJ}[
( - \frac{\partial V(\sigma_J,x_J)}{\partial x_J}  + \xi_{\sigma_J,
J})\: \mathbf{ \hat {d}}]
\end{displaymath}
and the analogous of eq. \ref{eq:1} is:
\begin{equation}
\dot{x}_I =\mathbf{v}_I \cdot \mathbf{ \hat {d}}
-\frac{1}{\gamma_{\sigma_I,I} }[\frac{\partial V(\sigma_I,x_I)}{\partial x_I}
+\xi_{\sigma_I, I}]\label{eq:4}
\end{equation}

The problem is hard to tackle analytically as is, because it is a 
selfconsistency problem in which the instantaneous configuration of the rowers 
affects the Stokes field, which in turn enters the equation of a single rower 
as the (local) velocity of the surrounding fluid.  
Nevertheless, a mean field calculation is fairly easy. That is, we
examine if a macroscopically steady constant flow can be sustained by
the beating rowers. \\  
We can write (see \cite{Doi}) the average velocity field in one point
\(\mathbf{R}\) of the surrounding fluid as

\begin{displaymath}
        \mathbf{v}(\mathbf{R}) =  \sum_{J} \bar{H}(\mathbf{R} - \mathbf{r}_{J}) 
        \gamma(\sigma_{J})( \mathbf{I}_{\sigma_J,J} - \mathbf{v})
\end{displaymath}
Which, averaged and projected along the direction of beating, taking
into account the average force exerted by the single rower, gives
the self-consistency relation for the constant fluid velocity  

\begin{displaymath}
        v_{\textrm{fluid}}  =\frac {H^{int} (\gamma_2 - \gamma_1)} {1
        + H^{int} (\gamma_2 + \gamma_1)} I(v_{\textrm{fluid}})
\end{displaymath}
The quantity \( {H}^{int} \) is a number that derives from the sum over the
(finite) lattice sites of the Oseen propagators, and \( I \) is 
the average active current, as defined in equation (\ref{eq:3}). 
If \( {H}^{int} \) is big the equation becomes
\( v_{\textrm{fluid}} = \frac {\gamma_2 - \gamma_1}{\gamma_2 + \gamma_1}
I(v_{\textrm{fluid}}) \), whereas in the limit of small \( {H}^{int}
\) we get \(v_{\textrm{fluid}} = H^{int} (\gamma_2 - \gamma_1)
I(v_{\textrm{fluid}}) \).  The value of \( {H^{int}} \) depends of
course on the arrangement of the rowers on the lattice, and can be
easily calculated. When the number of rowers is not finite problems
may arise because of the \( \frac{1}{r} \) dependence of the Oseen
tensor, but this situation is not realistic for systems of cilia.
Following Landau \cite{landau}, one could extend the sum over the
penetration length of the hydrodynamic interaction, so that the
constant will be a function of the surface density of rowers.\\
Then, the velocity field can be either zero, or take the (positive
or negative) selfconsistent value \( v_{SC} \ne 0 \) (fig. 3). \\
This means that the system of rowers is able to set up a macroscopic 
(and macroscopically steady) flow in the fluid. As this flow is
selfconsistently maintained by the array of rowers, we can see this
process as a spontaneous, dynamic symmetry breaking.
\begin{figure}[h]
\begin{center}
\includegraphics[scale=.63]{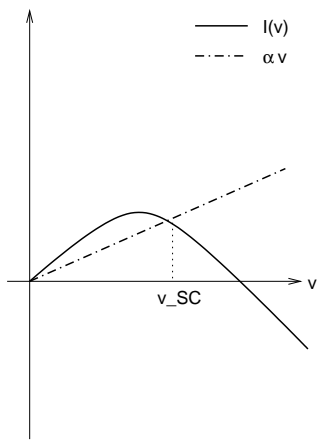}
\end{center}
\caption{Sketch of the selfconsistent velocity calculation. \( \alpha = \frac{1
        + H^{int} (\gamma_2 + \gamma_1)} {H^{int}(\gamma_2 - \gamma_1)} \) }
\label{fig:3} 
\end{figure}

The question whether this symmetry breaking process could be relevant
for real cilia is beyond the descriptive capabilities of the model.
Nevertheless, we have established that ideal, minimal, cilia-like
object as the rowers that are not intrinsically oriented, may achieve a
directionality collectively.


\section{Metachronal coordination}\label{sec.4}

The mean field approach of the above paragraph prevents by
construction the analysis of wave-like patterns in the beating of the
rowers. It tells us that the fluid is pumped by the rowers but not if
they pump it \emph{coordinately}.  \\
In this section we want to analyze the possible active role of
hydrodynamic interaction not only in breaking left-right symmetry
but also in creating patterns. 
Given that this interaction (alone) is able to generate directed fluid
flux, we are now looking for the premises for spatial coordination. 
To escape from the mean field description we turn to a path integral
representation of equation \ref{eq:4}. \\
The entity of interest is the effective action for the configuration
variables. It is from the minima and from the curvature of this action
that we expect to get evidence about the existence of waves
(patterns).

In our description we are considering the ``ground'' state of the
system to undergo continuous changes driven by the dynamics of the
discrete field \( \sigma \).  Thus, in our model the metachronal wave
cannot have the nature of a small oscillation around an equilibrium
state but rather it is a far-from-equilibrium oscillatory pattern.
This implies that standard techniques relying on conservation laws or
symmetry breaking (see for example \cite{forster}) cannot be employed
to find waves in the form of propagating modes.
In this view, the assumption that the time scales of thermalization
are fast compared to \( \sigma \) is very important in order to see 
any kind of oscillation.

In what follows we outline the calculation.

The reduced partition function to the configurational variables is
\begin{displaymath}
        Z({J_I(t)}) = \langle \exp[\sum_I \int \textrm{dt} J_I(t) x_I(t)] 
        \rangle_{\xi, \sigma}  
\end{displaymath}
The two averaging steps involved are integration on thermal noise \(
\xi\) and on the noise \( \sigma \).
Integration on thermal  noise is straightforward (see \cite{zinn-justin}).\\
The second integration is a much more delicate step.  First, \( \sigma
\) is not only an additive noise, but it has a multiplicative role
too, second \( \sigma \) has to be described on greater time scales
than the thermalization times.  

It is convenient to rewrite the mobility matrix as
\begin{displaymath}
        L_{IJ} = H_{IJ} + \eta \sigma_I \delta_{IJ}
\end{displaymath}
-- where the dependence on \( \sigma \) has been isolated and \( H_{IJ} \)
has the same off-diagonal terms as the Oseen tensor -- and to
approximate the inversion of \( L \) up to first order in \( \eta \).
We choose for simplicity to have the
potentials  \( V_{\sigma,I} = \frac{1}{2} a (x_I - \sigma_I)^2
\), so that the symmetry breaking is assumed and the two wells are 
quadratic with the same stiffness \( a \). 
This choice allows us to analyze the existence problem of the
metachronal wave using straightforward algebra. 
(Keeping into account more general potentials involves non-linearities 
which, however, do not affect space-time derivatives and consequently 
do not change our analysis about the metachronal wave.)\\
Integration on thermal noise is carried on with
\begin{displaymath}
  \langle\xi_I(t)\rangle=0 \quad ; \quad \langle\xi_I(t) \xi_J(t')\rangle = 
  L^{-1}_{IJ} \delta(t-t') 
\end{displaymath}
and gives an effective dynamical action depending on the fields \(x\) and 
\(\sigma\):
\begin{displaymath}
  S = \int \textrm{dt} ( \mathcal{L}_x +
  \mathcal{L}_{x,\sigma} + \mathcal{L}_{\sigma} ) 
\end{displaymath}

where \( \mathcal{L}_x \) depends only on the configurations,
 \(   \mathcal{L}_{\sigma} \) involves the field \( \sigma
\), and \(  \mathcal{L}_{x,\sigma}  \) is an interaction term.

We find
\begin{displaymath}
        \mathcal{L}_x =\sum_{I,J} x_I [ H^{-1} (- \partial^2_t +
        a^2 H^2 ) ] _{IJ} x_J +2 a^2 \eta \sum_I x_I  
\end{displaymath}
 \begin{displaymath}
        \mathcal{L}_{\sigma} = a^2  \sum_{I,J} \sigma_I H_{IJ} \sigma_J +
        \eta a^2 \sum_I  \sigma_I
\end{displaymath}
\begin{displaymath}
  \mathcal{L}_{x,\sigma} = 2 a \sum_I \sigma_I (\dot{x}_I + a\sum_J
  H_{IJ} x_J) 
  + \eta \sum_I \sigma_I [  a^2  x^2_I - 
  \sum_{PQ} H^{-1}_{IP} \dot{x}_P H^{-1}_{PQ} \dot{x}_Q ]
\end{displaymath}

Let us now consider the integration in \( \sigma \).\\
We proceed with M.S. technique and consider the functional measure 
for the integration in \( \sigma \).
The fact that such a noise has a cyclic nature 
cannot be ignored, so we write the measure in the form:
\begin{equation}
  d\Sigma(\sigma) = [d\sigma] \exp\{ -\frac{1}{2} \sum_{I,J} \int
  \textrm{dt}\: \sigma_I(t)\: \Sigma(\partial_t)_{IJ} \sigma_J(t) \} 
  \label{eq:misura} 
\end{equation} 

We can observe that, in view of this expression and the above ones,
the complete effective kernel for \( \sigma \) is 
\begin{equation}
  a^2 H + \Sigma
  \label{eq:acca}
\end{equation}
This quadratic form determines, through its lowest eigenvalue, the
most probable configuration around which one can study the
fluctuations. Assuming space-time translational invariance, the
spectrum will depend on a wave-vector \(k\) and a frequency \(\omega\).

The question whether the hydrodynamic interaction can give rise to
metachronism is then equivalent in this formalism to asking whether
adding \( H \) to \( \Sigma \) can change the minimum eigenvalue of
the quadratic form for the field \(\sigma\), determining a ground
state configuration corresponding to well defined wave-vector \( k^*\)
and frequency \( \omega^* \), both different from zero, that will 
generate the metachronal wave.\\

As we do not have a microscopic theory for the internal engines of the
rowers, we have some freedom to choose the probability measure in
formula \ref{eq:misura} \(\Sigma\). The case we would like to
investigate first is the one in which \( \sigma_I(t) \) are spatially
independent random variables. We will find that even in this case
there are wave like solutions, but they are canceled out by the noise
when averaging. 

We take a \(\Sigma\) which is diagonal in space and has kernel which
is \emph{not} monotonically increasing in time. This last
requirement is crucial.
In fact, with a noise with a monotonically increasing kernel 
the system should be purely dissipative.  It
is easy to prove that the equation for the classical (most probable) field  
admits the null-path as unique solution, with
fluctuations exponentially decreasing both in space and time.
In the large time limit the probability distribution becomes
\begin{displaymath}
        P_{\textrm{stat,hydr}} \sim e^{- \frac{1}{2} \sum_J x_J^2}
\end{displaymath}
so that \( H_{IJ} \) does not intervene in any way in the stationary
probability measure, but it may just modify the thermalization times.
This means that if one models the energy release with a stochastic 
``white-like'' process, the hydrodynamic interaction cannot have 
influence on the stationary configuration of the system, and cannot set 
up any spatial coordination.

Thus, we assume that \( \Sigma \) is
\emph{not} monotonically increasing and analyze this case.

Notice that what follows is independent of the specific choice of the
operator \( \Sigma \), for example one can write - as in the
Brazovskii model \cite{Brazovskii} -- \( \Sigma = (C^2 \: \partial^4_t
+ D^2\: \partial^2_t + B^2)\).  With this kernel the most favored
modulation in time is \( \omega^*= | \frac{D}{\sqrt{2}C}|\) and it can
be identified with the transition frequency. The fact that this
modulation may correspond to a zero--mode is not a problem
with a bounded field. 
\\
We proceed summing over \( \sigma \).  
Since in our case the 
zero--mode can not be resolved by exploiting symmetry, we must assume
that the field \( x_I(t) \) is limited in width.  The results that we
can easily obtain with this assumption are equivalent to the results
of a different, more heavy but mathematically more careful analysis.

Thus, we look for the classical solution with field \( x_I(t)
\) - the most probable path and first excitation
- of the form \( \underline x=\underline x^0+\eta \underline x^1 + \cdots \).

It is easy to verify that \(\underline x^0 \) e \(\underline x^1 \)
are solutions of the equations (from hereafter we will take \( a=1\)):

\begin{equation}
  \sum_{J}\big [(- \partial^2_t + H^2) \frac {\Sigma}{H(H+\Sigma)}
  \big ]_{IJ} \: x^0_J
  = 0 \label{eq:5}
\end{equation}

and 

\begin{equation}
  \sum_K \big [ (- \partial^2_t + H^2) \Sigma ]_{IK} x^1_K
  = 
  \sum_J H_{IJ}  - \frac{1}{2} \sum_J (- \partial_t + H )_{IJ} 
  [ (x^0)^2 + (H^{-1} \dot x^0)^2]_J  - \cdots \label{eq:6}
\end{equation}

Considering \( \underline x^0\) first, the positivity of \( (-
\partial^2_t + H^2)\) and the existence of \(H^{-1}\) imply (eq.
\ref{eq:5}) that \( \underline x^0\) is a solution if \( \Sigma \:
\underline x^0=0\). This means that the properties of the internal
state field \(\sigma \) are transferred to \( x \) trough functional
integration, so that the  classical path
is random in space. Moreover, it
describes a null velocity for the fluid environment 
(from the simmetry of \( H \)).

Consider now the first correction in \( \eta \).
As one can see by eq. \ref{eq:6}, \(\underline x^1 \) has a space-time
source. Nevertheless, like for \(\underline x^0\), there is no
explicit dependence on spatial variables.
It is relevant that, instead,
\( \underline x^1\) gives a non zero fluid velocity,
that is: the rowers are idoneous to pump fluid, but again without
coordination.

We shall now consider the fluctuations and limit our study to the
quadratic part of the effective action. Our task is to analyze the
paths which correspond to non-zero eigenvalues \( \lambda \) of the
operator in eq. \ref{eq:5} (which gives the solution \( \underline x^0\)). 
 
The paths are \(x_I(t)= e^ {i\:(\omega t+kI)}\) with dispersion relation
\begin{equation}
  \lambda = (\omega^2 + H(k)^2) 
  \frac{\Sigma (\omega) }{H(k) [H(k) +\Sigma(\omega )]} 
\end{equation}
For each \( \lambda \) this relation gives a ``band'' of solutions
with the same statistical weight.  Together with the true waves -- the
{\it metachronal waves} -- there is a solution of the same kind of \(
\underline x^0 \), namely with spatial {\it randomness}.  These
solutions are obtained, for each \( \lambda \), by considering the
limit \( k \to 0 \) and recalling that \( H \) is essentially the
inverse of the laplacian -- in fact, for \( k \to 0 \) the operator
that we are considering reduces to \(
\Sigma ( \omega )\).

This further level of analysis which includes fluctuations confirms
the mean field result on the presence of an effective macroscopic
pumping of the fluid by the rowers, as an effect which is first order
in \( \eta \). However, the hydrodynamic interaction is frustrated in
sustaining the metachronal waves because for every metachronal mode
there is a path, with the same probability, and the same random nature
of the classical solution.  Furthermore the metachronal waves are
always depressed with respect to the classical solution.
\\
In conclusion, without exchange of chemical information between
rowers, the sole hydrodynamic interaction does not generate
coordination. This is mainly due to the fact that, going back to
equation \ref{eq:acca}, if \( \Sigma \) is diagonal in space, adding
the term \( H \sim 1/k^2 \) does not determine a modification in the
minimum eigenvalue of the spectrum giving rise to a well-defined mode.  

The situation is different if the functional measure for the field \(
\sigma \) contains a spatial interaction, which can be short ranged,
between the internal states \( \sigma_I \). For example, one could
consider a nearest neighbor interaction with coupling constant
\(\alpha\), giving rise to a Laplacian on the lattice. This does not
affect the minimum around the homogeneous configuration in absence of
hydrodynamic interaction.

However, in presence of \(H\), the spectrum becomes
\begin{displaymath}
  H + \Sigma \sim \frac{1}{k^2} +\alpha  k^2 + \Sigma(\omega)  
\end{displaymath}
and there is a minimum for the particular value 
\( |k^*|^4= \frac{1}{\alpha} \) of the wave
vector, together with the usual value \( \omega^* \) for the
frequency.  Integrating on \( \sigma \) and looking for the
eigenvalues of the effective quadratic form for \( x \) one is forced
to keep this minimum energy ``spatio-temporal mode'' into account and
obtains wave-like solutions with frequency \( \omega^* \) and
wave-vector \( {k^*} \).

This solution can be called metachronal wave according to our
definition. It is sustained by hydrodynamic interaction, but it needs
a preexisting short ranged interaction between the internal states of
the rowers to be formed. This preexisting interaction is unable by
itself to set up a mode. 

The physical interpretation for this short ranged coupling could be
that one cilium can feel the depletion in ATP concentration due to the
activity of nearby cilia of the same cell.  
\\
In order to obtain metachronism one can also consider an alternative
scenario in which the \(\sigma\) is dynamically related to the
configuration.  This scenario includes as a special case the
stochastic analogous of the ``geometric switch'' mechanism found in
\cite{gueron2} and \cite{gueron1}, in which the transitions between
the active and passive phases of the cilium are determined by its
reaching some limit configurations.

If we include a dependence on the configuration in the dynamical
equation for the field \( \sigma\),
so that the quantities \( \omega_i\) in eq. \ref{eq:2}
become necessarily functions of the space coordinate 
of the rower. The results of sections 2 and 3 do not change.
On the contrary, the functional integral study undergoes a dramatic 
change. Time modulation of \(\sigma\) noise need not be 
required \emph {ab initio}.
In this case there are two interacting fields, \( x_I(t) \) and \(
\sigma_I(t) \), the functional integral is well defined 
and a correct perturbative analysis can be carried on.
The scenario is formally equivalent to the one for modulated phases 
in membranes with defects \cite{leiblerfasi}. 
Dealing with quadratic potentials, for example,  
it is possible to integrate out the continuous configurational degrees of
freedom --  as in \cite{bassettix} --  obtaining an
effective model for the field \( \sigma\). 


\section{Conclusions} \label{sec.5}

We introduced a model system, the rower, which contains some
essentials of the cilium and, being economic in degrees of
freedom, enables to deal with stochastic features of this system. 

We computed the probability current of one rower interacting with a
surrounding fluid in a steady state, and we used the result to deal
with the problem of left-right symmetry breaking of this entity. 

The same expression of the current was then used in a self-consistent
mean field-like calculation for a planar array of rowers coupled
hydrodynamically. The result was that rowers can cooperate to set up a
macroscopic flow in the fluid.

Finally, we presented the problem of metachronal coordination in terms
of correlation between rowers, and discussed a path-integral
calculation that enables to point out some features that are
sufficient for the model to exhibit this behavior. 

This kind of calculation can be a useful tool in general for systems
driven far from equilibrium by a stochastic process that switches the
Hamiltonian locally.

The indications that come from the last two calculations are that
\begin{itemize}
\item{1)} For our rowers the metachronal wave is not necessary to set up a
  macroscopic flow in the surrounding fluid. This is supported by
  a mean field like analysis and confirmed when we include fluctuation.
  
\item{2)} Without any direct interaction between rowers the
  hydrodynamic interaction generates metachronal waves which are
  frustrated by the presence of random fluctuations of the same
  statistical weight, together with the random dominant solution.
 
\item{3)} A short ranged coupling of internal states (that could have
  for example chemical origin), unable by itself to set up a mode, can
  stabilize the wave and make the pattern formation statistically
  relevant.
  
\item{4)} Alternatively, provided that the only interaction between
  rowers is hydrodynamic, a sufficient condition for the onset of a
  metachronal wave is the presence of a coupling between position and
  transition frequency of the single rower.

\end{itemize}
 
These results are qualitative theoretical predictions. They have a
definite interest from the point of view of the model, but they need
to be examined in greater detail to fully understand their
implications for the real system.


\end{document}